\documentclass[%
 aip,
 mph,%
 amsmath,amssymb,
 superscriptaddress,
 floatfix,
preprint,
eqsecnum,%
numerical,%
]{revtex4-1}

\usepackage{graphicx}
\DeclareGraphicsExtensions{.eps,.pdf,.png,.jpg}
\usepackage[caption=false
]{subfig}
\usepackage{epsfig}
\usepackage{epstopdf}
\usepackage{lipsum}
\usepackage{wrapfig}
\usepackage{makecell}
\usepackage{tabularx}
\usepackage{multirow}
\usepackage{booktabs}
\usepackage{lmodern}
\usepackage{longtable}
\usepackage{amsmath}
\usepackage[OT2,T1]{fontenc}
\usepackage[bookmarks=false]{hyperref}
\usepackage{color}
\usepackage{float}
\hypersetup{pdfauthor={},pdfborder=0 0 0,colorlinks=true,citecolor=blue,linkcolor=blue,urlcolor=blue,}

\DeclareSymbolFont{cyrletters}{OT2}{wncyr}{m}{n}
\DeclareMathSymbol{\comb}{\mathalpha}{cyrletters}{"58}

\newcommand{\ben}{\begin{eqnarray}\displaystyle}
\newcommand{\een}{\end{eqnarray}}

\begin{document}

\title{Super resolution dual-layer CBCT imaging with model-guided deep learning}

\author{Jiongtao Zhu}
\thanks{Jiongtao Zhu and Ting Su have made equal contributions to this work and are considered as the first authors.}
\affiliation{Key Laboratory of Optoelectronic Devices and Systems of Ministry of Education and Guangdong Province, College of Physics and Optoelectronic Engineering, Shenzhen University, Shenzhen 518060, China}
\author{Ting Su}
\thanks{Jiongtao Zhu and Ting Su have made equal contributions to this work and are considered as the first authors.}
\author{Xin Zhang}
\author{Han Cui}
\author{Yuhang Tan}
\affiliation{Research Center for Medical Artificial Intelligence, Shenzhen Institute of Advanced Technology, Chinese Academy of Sciences, Shenzhen, Guangdong 518055, China}


\author{Hairong Zheng}
\affiliation{Paul C Lauterbur Research Center for Biomedical Imaging, Shenzhen Institute of Advanced Technology, Chinese Academy of Sciences, Shenzhen, Guangdong 518055, China}

\author{Dong Liang}
\affiliation{Research Center for Medical Artificial Intelligence, Shenzhen Institute of Advanced Technology, Chinese Academy of Sciences, Shenzhen, Guangdong 518055, China}
\affiliation{Paul C Lauterbur Research Center for Biomedical Imaging, Shenzhen Institute of Advanced Technology, Chinese Academy of Sciences, Shenzhen, Guangdong 518055, China}

\author{Jinchuan Guo}
\affiliation{Key Laboratory of Optoelectronic Devices and Systems of Ministry of Education and Guangdong Province, College of Physics and Optoelectronic Engineering, Shenzhen University, Shenzhen 518060, China}

\author{Yongshuai Ge}
 \thanks{Scientific correspondence should be addressed to Yongshuai Ge (ys.ge@siat.ac.cn).}
\affiliation{Research Center for Medical Artificial Intelligence, Shenzhen Institute of Advanced Technology, Chinese Academy of Sciences, Shenzhen, Guangdong 518055, China}
\affiliation{Paul C Lauterbur Research Center for Biomedical Imaging, Shenzhen Institute of Advanced Technology, Chinese Academy of Sciences, Shenzhen, Guangdong 518055, China}

\date{\today}

\begin{abstract}
{\bf Objective:} This study aims at investigating a novel super resolution CBCT imaging technique with the dual-layer flat panel detector (DL-FPD).\\
\noindent
{\bf Approach:} In DL-FPD based CBCT imaging, the low-energy and high-energy projections acquired from the top and bottom detector layers contain intrinsically mismatched spatial information, from which super resolution CBCT images can be generated. To explain, a simple mathematical model is established according to the signal formation procedure in DL-FPD. Next, a dedicated recurrent neural network (RNN), named as suRi-Net, is designed by referring to the above imaging model to retrieve the high resolution dual-energy information. Different phantom experiments are conducted to validate the performance of this newly developed super resolution CBCT imaging method.\\
\noindent
{\bf Main Results:} Results show that the proposed suRi-Net can retrieve high spatial resolution information accurately from the low-energy and high-energy projections having lower spatial resolution. Quantitatively, the spatial resolution of the reconstructed CBCT images of the top and bottom detector layers is increased by about 45\% and 54\%, respectively.\\
\noindent
{\bf Significance:} In future, suRi-Net provides a new approach to achieve high spatial resolution dual-energy imaging in DL-FPD based CBCT systems.\\
\end{abstract}


\maketitle

\section{INTRODUCTION}
\label{sec:introduction}
Recently, X-ray flat panel detector (FPD) with more than one single receptor, i.e., CsI:TI scintillator layer, begins to attract a lot of research interests in medical cone-beam CT (CBCT) imaging applications\cite{schulze2020cone,Oldham2005ConebeamCTGR,Wallace2008ThreedimensionalCC}. As the number of receptor increases, it becomes possible for FPD to differentiate the incident X-ray photons by their energies. Take the dual-layer FPD (DL-FPD) as an example, X-ray photons carrying lower energies are mostly absorbed by the top layer, and the rest X-ray photons carrying higher energies are mostly absorbed by the bottom layer. As a result, DL-FPD based CBCT imaging\cite{shi2020characterization} can generate temporally registered spectral-spatial information, i.e., dual-energy CBCT projections, from which material-specific bases that are valuable for disease diagnoses\cite{gang2012cascaded, patino2016material, nicolaou2010dual} can be obtained.

By far, the performance of DL-FPD based dual-energy CBCT imaging has been investigated by several studies. For instance, Shi \textit{et al.} and St{\aa}hl \textit{et al.} demonstrated the feasibility of performing accurate dual-energy CBCT imaging based on DL-FPD \cite{shi2020characterization, staahl2021performance}. Wang \textit{et al.} proposed a model-based high resolution material decomposition method for DL-FPD based CBCT\cite{Wang2021HighresolutionMM}. Despite of its feasibility, DL-FPD may have several limitations in generating high-quality dual-energy CT basis images. For example, spectra separation maybe less satisfied in DL-FPD. To compensate, additional beam filtration is often inserted between the top and bottom receptors to increase the energy separation\cite{shi2020characterization}. However, the radiation dose efficiency might be degraded. In addition, the spatial resolution performance, which is usually characterized by the modulation transfer function (MTF), of the top and bottom receptors is different. Often, the CsI:TI scintillator is made thicker on the bottom receptor to efficiently stop the high energy X-ray photons. Hence, MTF of the bottom receptor is lower\cite{lu2019dual} than that of the top receptor made with thinner CsI:TI scintillator. Finally, the spatial resolution of the CBCT images may be degraded due to the detector element binning. In order to speed up the data acquisition procedure, the intrinsic receptor array is usually binned into a smaller array size at the expense of spatial resolution loss. Take the PaxScan 4030CB FPD (Varex, USA) as an example, X-ray signals captured by the native 2048 $\times$ 1536 detector elements with dimension of 0.194~mm (limiting resolution 2.58 lp/mm) can only be accessed by 7.5 fps per second at most\cite{WinNT,sheth2022technical}. As a contrary, the $2\times2$ binned 1024 $\times$ 768 detector elements with dimension of 0.388~mm (limiting resolution 1.29 lp/mm) can capture the X-ray signals at a speed of 30.0 fps. Apparently, detector binning would greatly accelerate the signal readout speed in FPD and DL-FPD, but also would dramatically degrade the spatial resolution. 

To improve the spatial resolution of CT images, a series of innovative methods have been proposed over the past two decades. For example, Tilley \textit{et al.} proposed to enhance the image resolution via model-based iterative reconstruction (MBIR) that incorporates system blur and noise correlation \cite{Tilley2016ModelbasedIR, Tilley2018PenalizedLikelihoodRW}. Hashemi \textit{et al.} proposed a simultaneous deblurring and iterative reconstructing method for high resolution CBCT imaging\cite{Hashemi2017SimultaneousDA}. Aarle \textit{et al.} proposed a super-resolution CT image reconstruction approach based on prior knowledge of the imaging object \cite{Aarle2014SuperResolutionFC}. Furthermore, deep-learning based super resolution CT imaging methods have also been proposed recently. You \textit{et al.} developed a semi-supervised residual CycleGAN network to generate high-resolution CT images from low-resolution CT images\cite{You2020CTSG}. Chen \textit{et al.} proposed a network-based iterative CBCT reconstruction method to learn the regularization term and enhance the image resolution\cite{Chen2018StatisticalIC}. Hatvani \textit{et al.} used a modified U-net and a subpixel network to map the low-resolution CBCT images onto the high-resolution micro-CT images\cite{Hatvani2019DeepLS}. Besides the aforementioned innovative CT image reconstruction approaches, essentially, the CT image spatial resolution can also be enhanced with advanced X-ray hardware such as high-end X-ray source and detector. On the X-ray source end, the flying focal spot (FFS) technique \cite{Kachelriess2006FlyingFS,Flohr2005ImageRA,kyriakou2006impact} is able to double the spatial sampling density in the horizontal and vertical directions. On the X-ray detector end, sub-pixel shifting\cite{Yan2015SuperRI,Li2020MicroCTIO} is usually implemented. For instance, a super resolution CBCT imaging strategy\cite{su2022super} incorporating additional 1D or 2D sub-pixel shift between the top and bottom receptors of DL-FPD is proposed.

\begin{figure*}[htb]
\centerline{\includegraphics[width=1.0\textwidth]{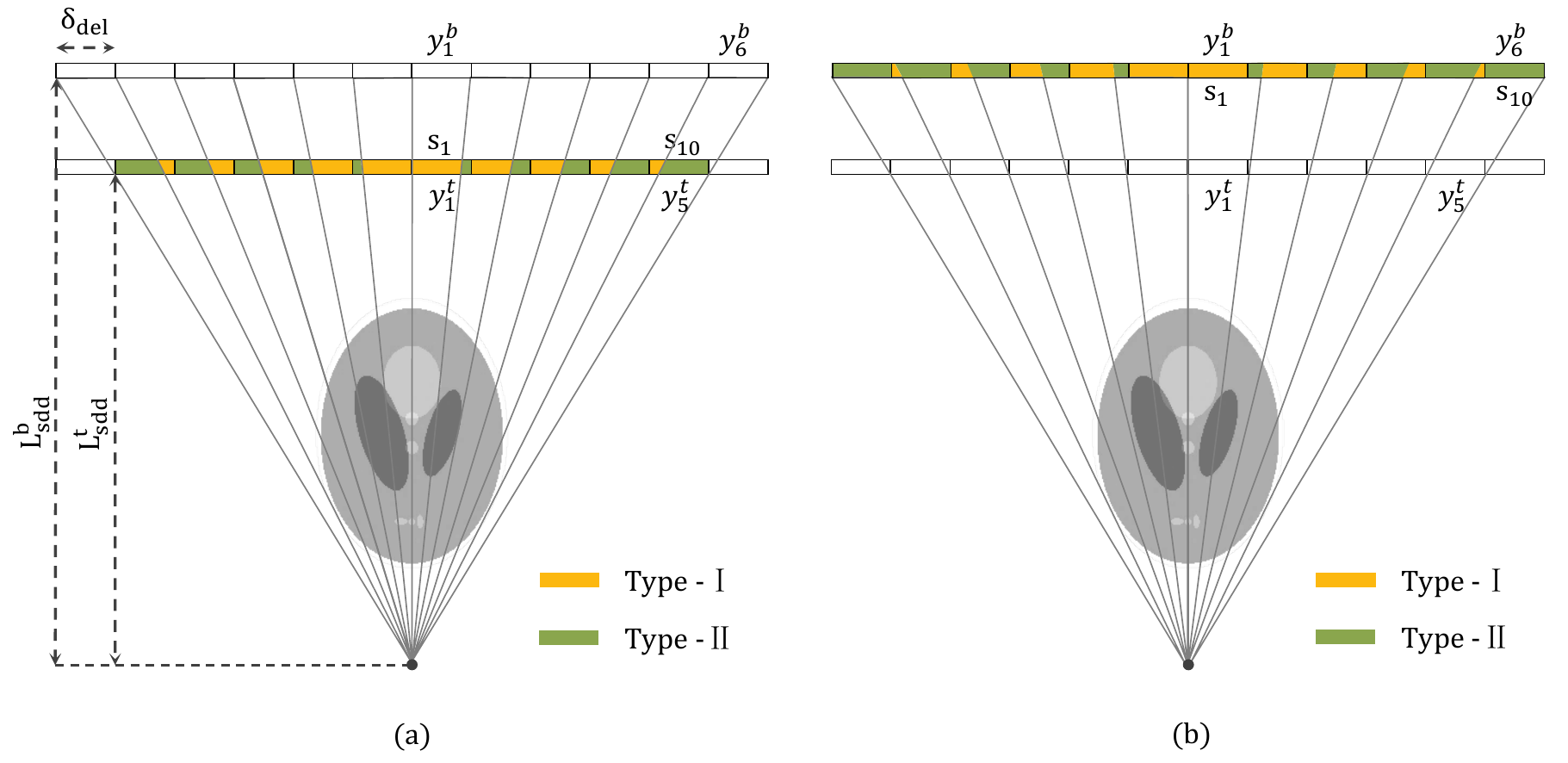}}
\caption{Illustrations of the formation of sub-pixels (colored segments with varied widths) in DL-FPD imaging. (a) The sub-pixels on the top layer, and the reference is the bottom layer. (b) The sub-pixels on the bottom layer, and the reference is the top layer.}
\label{Imaging_model}
\end{figure*}

Herein, an alternative super resolution CBCT imaging method is suggested for DL-FPD. Instead of introducing additional 1D or 2D sub-pixel shift between the top and bottom layers, this new approach relies on the intrinsically mismatched spatial information between the top and bottom detector layers to generate super resolution CBCT images. Usually, there exists a small gap between the top and bottom layers in DL-FPD. As a result, mismatched spatial information can be acquired from the low-energy and high-energy projections, see Fig.~\ref{Imaging_model}. As illustrated, a certain beamlet carrying the same integration information fall on two different locations of the top and bottom receptor planes. Therefore, it is possible to restore the spatial information at sub-pixel scale, i.e., super resolution imaging, from such spatially mismatched dual-layer projections. To do so, a dedicated recurrent neural network (RNN), denoted as suRi-Net, is designed according to the signal formation procedure to retrieve such high spatial resolution projections. Afterwards, material-specific basis images with high spatial resolution are reconstructed via the image-domain decomposition algorithm.

The rest of this paper is organized as follows: Section \ref{sec: method} presents the mathematical model, assumptions, the SuRi-Net network, experimental setup and implementation details. Section \ref{sec:results} presents the experimental results of different phantoms. Section \ref{sec:conclusion} provides the discussions and a brief conclusion. 

\section{METHOD} \label{sec: method}
\subsection{Imaging model}

\begin{figure*}[!t]
\centerline{\includegraphics[width=0.9\linewidth]{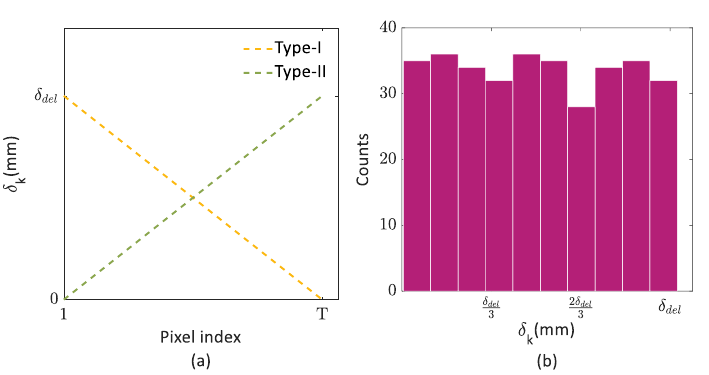}}
\caption{(a) Plots of $\delta_k$ for Type-I and Type-II sub-pixels. (b) The distribution of $\{\delta_k\}$.}
\label{subpixel_size}
\end{figure*}

Assuming the detected signals (proportional to the total number of X-ray photons) on the top and bottom layers are denoted as $y^{t}_{j}$ and $y^{b}_{i}$, correspondingly. With respect to a certain detector element $y^{b}_{i_0}$ on the bottom layer, see Fig.~\ref{Imaging_model}(a), the coverage (intersection length) of the same X-ray beamlet on the top layer is equal to:
\begin{equation}
\delta'_{del} = \frac{L^{t}_{sdd}}{L^{b}_{sdd}} \times \delta_{del} < \delta_{del}, \\
\label{eq0}
\end{equation}
where $L_{sdd}$ denotes the source to detector distance (SDD), and $\delta_{del}$ denotes the dimension of a detector element. Due to the fan-beam imaging geometry, obviously, the transmission information collected in $y^{t}_{j=m}$ and $y^{b}_{i=m}$ are not identical. Quantitatively, 
\begin{equation}
\begin{cases}
y_{1}^{b}=s_{1}, \\
y_{1}^{t}=s_{1}+s_{2}, \\
y_{2}^{b} = s_{2}+s_{3}, \\
y_{2}^{t} = s_{3}+s_{4}, \\
...
\end{cases} 
\label{eq1}
\end{equation}
Herein, monochromatic X-ray beams are assumed, and $s_{k}$ denotes the signal intensity of a sub-pixel. As a consequence, this provides an opportunity to reconstruct X-ray projections with finer spatial resolution, see the colored sub-pixels in Fig.~\ref{Imaging_model}. Analytically, 
\begin{equation}
\begin{cases}
s_{1}=y_{1}^{b}, \\
s_{2}=y_{1}^{t}-y_{1}^{b}, \\
s_{3}=(y_{1}^{b}+y_{2}^{b})-y_{1}^{t}, \\
s_{4}=(y_{1}^{t}+y_{2}^{t})-(y_{1}^{b}+y_{2}^{b}), \\
...
\end{cases} 
\label{eq2}
\end{equation}
In fact, the sub-pixel signals in Eq.~(\ref{eq2}) can be generalized into two main types. For the Type-I sub-pixel, it has
\begin{equation}
s_{2i_0-1} = 
\begin{cases}
y^{b}_{i_0}, \;\;\;\ i_0=1\\
{\textstyle \sum_{i=1}^{i_0}}y^{b}_{i}-{\textstyle \sum_{j=1}^{i_0-1}}y^{t}_{j}, \;\;\;\ i_0>1
\end{cases} 
\label{eq3}
\end{equation}
For the Type-II sub-pixel, it has
\begin{equation}
s_{2i_0} = {\textstyle \sum_{j=1}^{i_0}}y^{t}_{j}-{\textstyle \sum_{i=1}^{i_0}}y^{b}_{i}, \;\;\;\ i_0>1
\label{eq4}
\end{equation}
Notice that the above results are derived for the detector elements on the bottom layer with index $1\le i_0\le N/2$. Similar expressions can also be derived for the detector elements on the bottom layer with index $-N/2 \le i_0\le-1$, and so for the depicted scenario in Fig.~\ref{Imaging_model}(b). Interestingly, the width of a sub-pixel, denoted as $\delta_{k}$, varies from one to the another. Specifically, the width $\delta_{k}$ of Type-I sub-pixel is equal to
\begin{equation}
\delta_{2i_0-1} = 
\begin{cases}
\delta'_{del}, \;\;\;\ i_0=1\\
i_0\times\delta_{del}-(i_0-1)\times\delta'_{del}, \;\;\;\ i_0>1 
\end{cases}
\label{eq5}
\end{equation}
And the width $\delta_{k}$ of Type-II sub-pixel is equal to
\begin{equation}
\delta_{2i_0} = i_0\times(\delta_{del}-\delta'_{del}), \;\;\;\ i_0>1
\label{eq6}
\end{equation}
The relationship between $\delta_{k}$ and $i_0$ is plotted in Fig.~\ref{subpixel_size}(a), and the distribution of $\{\delta_k\}$ is plotted in Fig.~\ref{subpixel_size}(b). In the above derivations, it is assumed that
\begin{equation}
i^{max}_0 \le T = \lfloor \frac{\delta_{del}}{\delta_{del}-\delta'_{del}}\rfloor.
\label{eq7}
\end{equation}
As a result, the maximum detector dimension is assumed equal to $2T \times \delta_{del}$ in this study. Be aware that similar results can also be derived for detectors having larger dimensions.

By far, we have mathematically demonstrated the capability of obtaining higher resolution X-ray projections from the DL-PFD. Despite of its simplicity, unfortunately, Eq.~(\ref{eq3}) and Eq.~(\ref{eq4}) are quite difficult to be solved with a practical DL-FPD imaging system due to several non-idealities: First, the attenuation information carried by the low-energy and high-energy X-ray beams are different. As a result, the dual-layer projections can not be subtracted directly to generate valuable (quantitatively faithful) signal intensities. Second, the derived analytical solutions in Eq.~(\ref{eq3}) and Eq.~(\ref{eq4}) may introduce artifacts when image noise is consideried. Third, it is not very straightforward to reconstruct high resolution CT images from the estimated projections with dramatically varied sub-pixels dimensions.

\begin{figure*}[t]
\centerline{\includegraphics[width=1.0\linewidth]{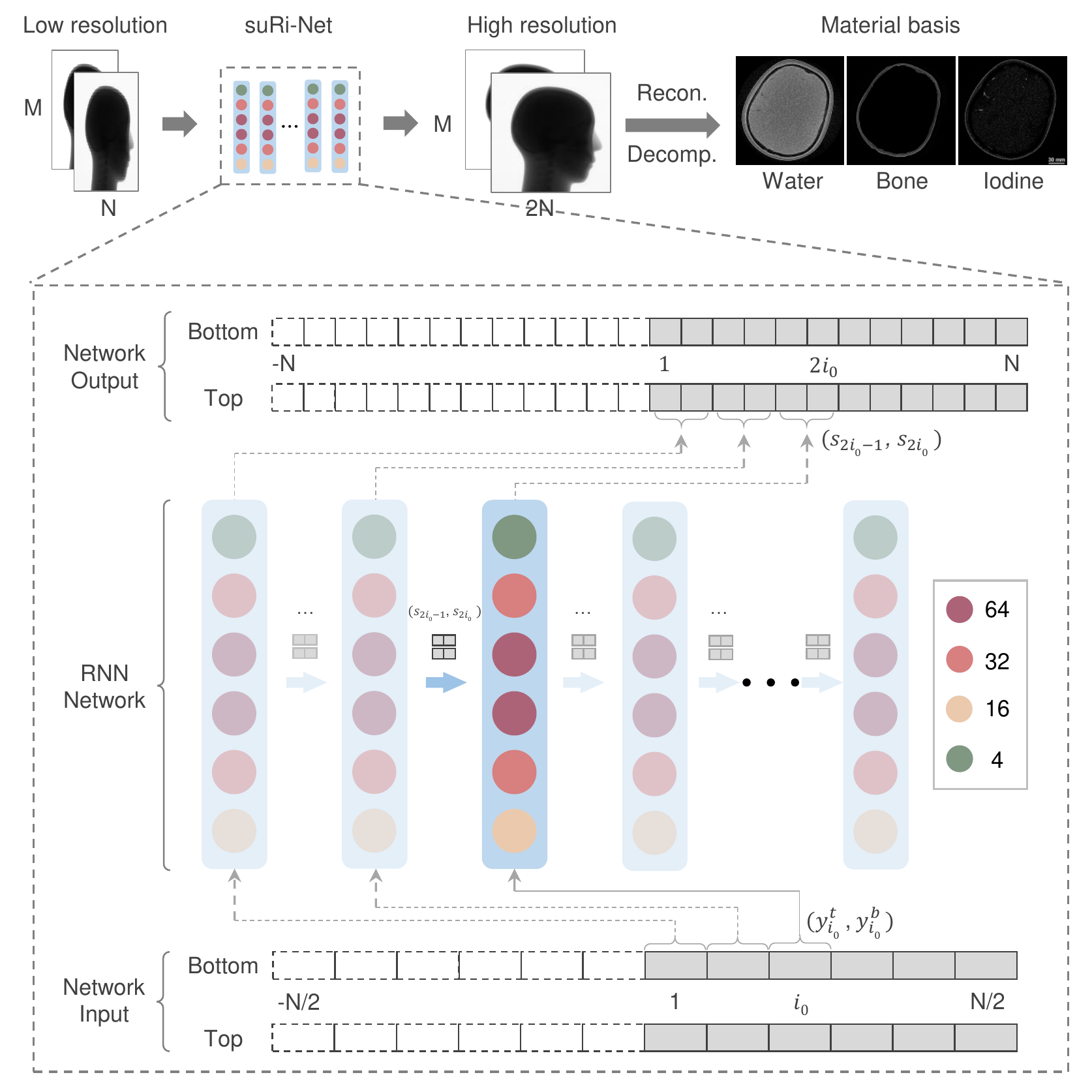}}
\caption{The process of dual-energy super resolution imaging and the structure of suRi-Net. The number of hidden units in each RNN layer is 16, 32, 64, 64, 32, and 4, respectively.}
\label{Network_structure}
\end{figure*}

\subsection{suRi-Net}
To overcome the above difficulties encountered by Eq.~(\ref{eq3}) and Eq.~(\ref{eq4}), the following two assumptions are made: First, the element dimensions of the high resolution low-energy and high-energy projections are identical: $\delta_k=\langle\delta_{del}\rangle = 0.5\delta_{del}$. This assumption makes sense only in a statistical way, see the nearly uniform distribution of $\delta_k$ between 0 and $\delta_{del}$ in Fig.~\ref{subpixel_size}(b). Second, the signal retrieval procedures defined in Eq.~(\ref{eq3}) and Eq.~(\ref{eq4}) can be re-formulated as follows,
\begin{equation}
(S^t, S^b)=\mathcal{R}(Y^t, Y^b),
\label{eq11}
\end{equation}
where symbols $S$ and $Y$ denote the super resolution projections and the low resolution projections, correspondingly, $\mathcal{R}$ denotes the complicated non-linear signal extraction operation.

In Eq.~(\ref{eq3}) and Eq.~(\ref{eq4}), it demonstrates that the computation of $s_{2i_0-1}$ relies on the accumulated summation of all preceding pixels before $y^{b}_{i_0}$ and $y^{t}_{i_0-1}$. Similarly, the computation of $s_{2i_0}$ relies on the accumulated summation of all preceding pixels before $y^{b}_{i_0}$ and $y^{t}_{i_0}$. In short, the computation procedure of $s_{k}$ has a sequential dependency on $y^{b}_{j}$ and $y^{t}_{i}$. Inspired by such sequential dependency and calculations, the Recurrent Neural Network (RNN) is employed to solve Eq.~(\ref{eq11}). As known, RNN is a type of artificial neural network good at handling sequential calculations\cite{lipton2015critical}, which allows previous outputs to be used as inputs while having hidden states. Therefore, it is considered suitable and capable to retrieve the desired super resolution projections with increased accuracy and efficiency.

In this study, such a dedicated RNN network is named as suRi-Net, see Fig.~\ref{Network_structure}. Specifically, the suRi-Net contains six hidden layers to extract image information, and each hidden layer uses the Leaky-ReLU as the activation function. A pair of detector responses $(y^t_{j=i_0}, y^b_{i=i_0})$ are sequentially put into the $i^{th}_0$ RNN layer together with outputs from the previous layer are fed into the $i^{th}_0$ RNN layer. The input low-energy and high-energy projections have sizes of M$\times$N, while the output projections have sizes of M$\times$2N. After being processed by the RNN, low-energy and high-energy signals with higher spatial resolution can be obtained, see Fig.~\ref{Network_structure}.

The mean squared error (MSE) between the network predicted high resolution dual-layer projections $\vec S=(S^t, S^b)$ and the label $\vec S_0=(S^t_0, S^b_0)$ is used as the network loss:

\begin{equation}
\mathcal L =\frac{1}{\rm{M} \times \rm{N} \times 2} \left\| \vec S - \vec S_0 \right \|_2, \\
\label{eq12}
\end{equation}

where $\left \| \cdot \right \|_2$ denotes the $l_2$ norm.

The training of suRi-Net was performed on TensorFlow platform on a workstation with Nvidia Titan RTX GPU. The initial learning rate was set to $1\times10^{-4} $, and decayed by 0.98 in every 3000 steps. The network was trained for 350,000 steps with a batch size of 1 for 720,000 pairs of one-dimentional training data. 

\subsection{Data preparation}
The training data are numerically simulated\cite{zhu2022feasibility, Su2021DIRECTNetAU} in Python with the images obtained from ImageNet\cite{deng2009imagenet}. The imaging geometry is set as: $L^{t}_{sdd} = 1300$ mm, $L^{b}_{sdd} = 1306.6$ mm, $\delta_{del} = 0.616$ mm and $\delta_{k} = 0.308$ mm. Specifically, the high resolution ($\delta_{k} = 0.308$ mm) labels are simulated at first. Afterwards, the low resolution projections ($\delta_{del} = 0.616$ mm) are obtained from them via 1$\times$2 binning. The low-energy and high-energy X-ray beam spectra were obtained as follows: the incident beam spectra was simulated at 125 kVp with 1.5 mm aluminum(Al) and 0.4 mm Copper(Cu) filtration in SpekCale\cite{Poludniowski_2009}, the low-energy beam spectrum corresponds to the absorbed spectrum on the 0.26 mm thick CsI:TI scintillator layer, and the high-energy beam spectrum is calculated by adding additional 1.0 mm Cu filtration and 0.26 mm CsI:TI scintillator layer. The number of X-ray photons is set to $1\times10^6$ per detector element, and Poisson noise is added.

\begin{table}[htb]
\caption{Key parameters used in experiments.}
\label{table1}
\centering	
\begin{tabular}{p{200pt}p{200pt}}
\hline\hline
Parameter            & Value\\ \hline
Tube voltage (kVp)        & 125                                                                    \\ 
Tube Current (mA)    & 12.5                                                                   \\ 
Output filtration (mm)     & Al: 1.5, Cu: 0.4                                                             \\ 
Total views            & 450 \\ 
SOD (mm)             & 1160.0                                                                 \\ 
SDD (mm)             & 1300.0 (top), 1306.6 (bottom) \\ 
Image matrix          & Limestone: 800$\times$800 \\ & CatPhan: 600$\times$600 \\ & KYOTO: 512$\times$512\\ 
Voxel size (mm$^3$)       & Limestone: 0.2$\times$0.2$\times$0.2 \\ & CatPhan: 0.35$\times$0.35$\times$0.35 \\ & KYOTO: 0.4$\times$0.4$\times$0.4 \\ 
Detector array        & 768$\times$768                                                              \\ 
Pixel size (mm$^2$)    & 0.308$\times$0.308                                                                  \\ 
Frame rate (fps)    & 7.5 \\ 
Manual binning       & 1$\times$2                                                                    \\ 
CsI:TI thickness (mm)            & 0.26 (top), 0.55 (bottom)\\ \hline\hline
\end{tabular}
\end{table}

\subsection{Experiments}
Physical experiments are performed on an in-house dual-layer CBCT imaging benchtop. The rotating-anode X-ray tube (Varex G-242, USA) is operated at 125 kVp. The DL-FPD detector (560RF-DE, CareRay, China) has a 768$\times$768 array and 0.308 mm$\times$0.308 mm element dimension (2$\times$2 binned). The CsI:TI scintillator in the top layer is 0.26 mm thick, and is 0.55 mm in the bottom layer. The two layers is 6.6 mm apart, and an additional 1.0 mm Cu is inserted in between to harden the high-energy X-ray beam. In addition, a 0.4 mm Cu is added to filter the incident X-ray beam. To minimize the Compton scattering effect, X-ray beam with 15.0 mm width (on the detector surface) is collimated. In total, 450 projections are collected from a full 360$^{\circ}$ rotation with 0.8$^{\circ}$ angular interval. Three different objects were scanned: porous limestone rods with a diameter of 65.0 mm, Catphan-700 phantom (CTP682 and CTP714 modules, Phantom lab, USA), and a head phantom (41309-300, Kyoto Kagaku, Japan), see more details in Table \ref{table1}.

Be aware that these originally acquired projections from this DL-FPD is used to reconstruct the ground truth CT images of high resolution. To generate the low resolution projections, these high resolution projections are manually $1\times2$ binned along the horizontal direction. Unless specified, the binning mode mentioned in the rest of the manuscript refers to the manual binning. For example, $1\times1$ binning denotes the high resolution projections, $1\times2$ binning denotes the low resolution projections.

\section{Results}\label{sec:results}

\begin{figure*}[htb]
\centerline{\includegraphics[width=0.9\linewidth]{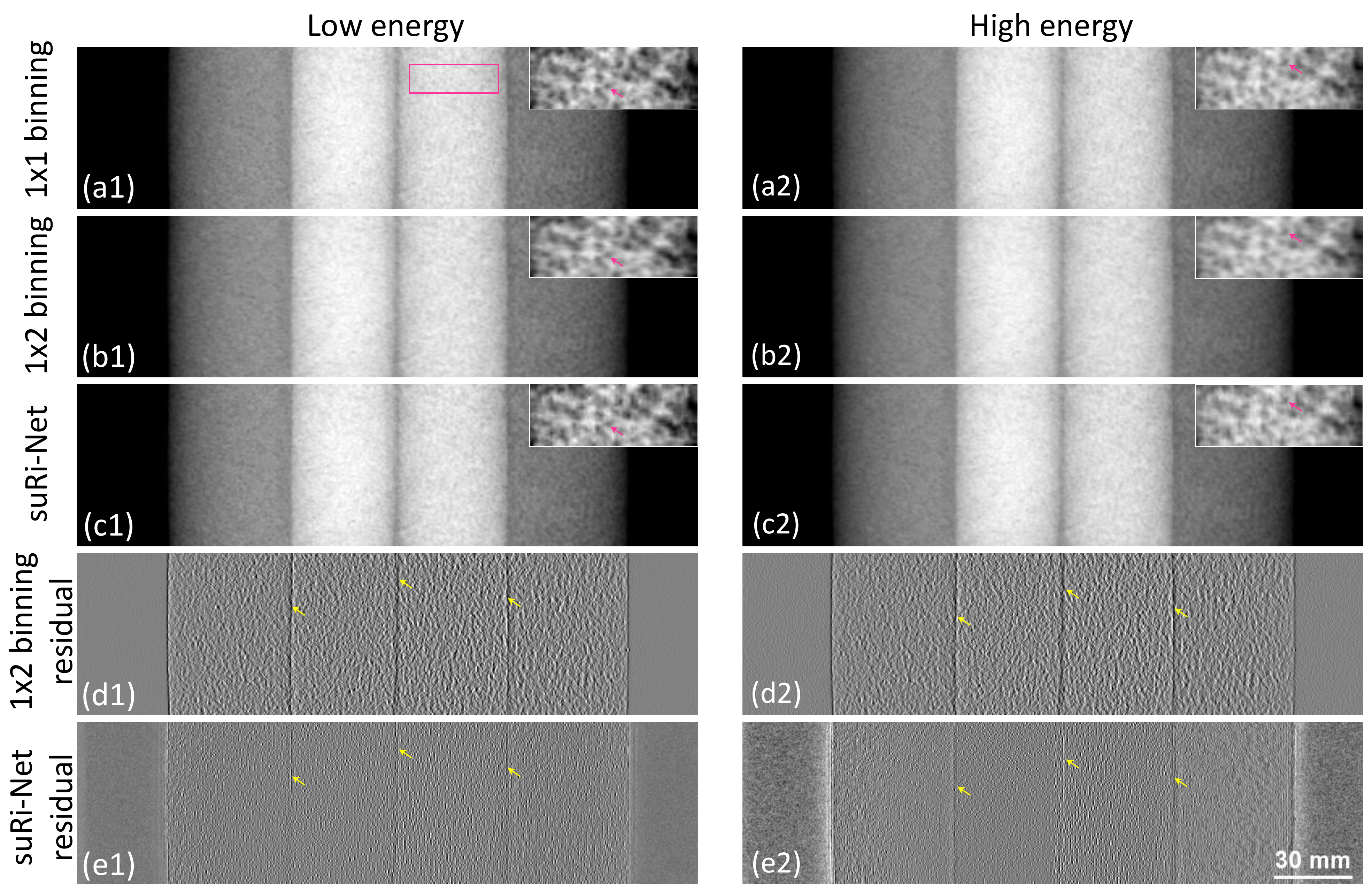}}
\caption{Projection results of the limestone rods. The reference projections with 1x1 binning are depicted in the first row, the 1x2 binned projections are depicted in the second row, and the projections obtained from suRi-Net are depicted in the third row. The residual differences are depicted in the fourth and fifth rows. The display windows for the low-energy and high-energy projections are [-0.0018, 2.4916] and [0.0287, 2.021], respectively. The display windows for the residual images are [-0.0425, 0.0407] and [-0.0290, 0.0277], respectively.}
\label{Limestone-p}
\end{figure*}

The projections of the porous limestone rods are shown in Fig.~\ref{Limestone-p}. In it, the high resolution reference projections with 1$\times$1 binning are presented in the first row, the low resolution projections with 1$\times$2 binning are presented in the second row, and the generated high resolution projections from the proposed suRi-Net are presented in the third row. As seen from the zoomed-in region-of-interest (ROI), results obtained from suRi-Net can greatly improve the spatial resolution for both low-energy and high-energy projections. Such improvements can be more easily observed from the subtracted residual difference images, see the highlighted arrows in the fourth and fifth rows. Herein, the residual differences are obtained with respect to the high resolution projections of 1$\times$1 binning.

\begin{figure*}[htb]
\centerline{\includegraphics[width=1\linewidth]{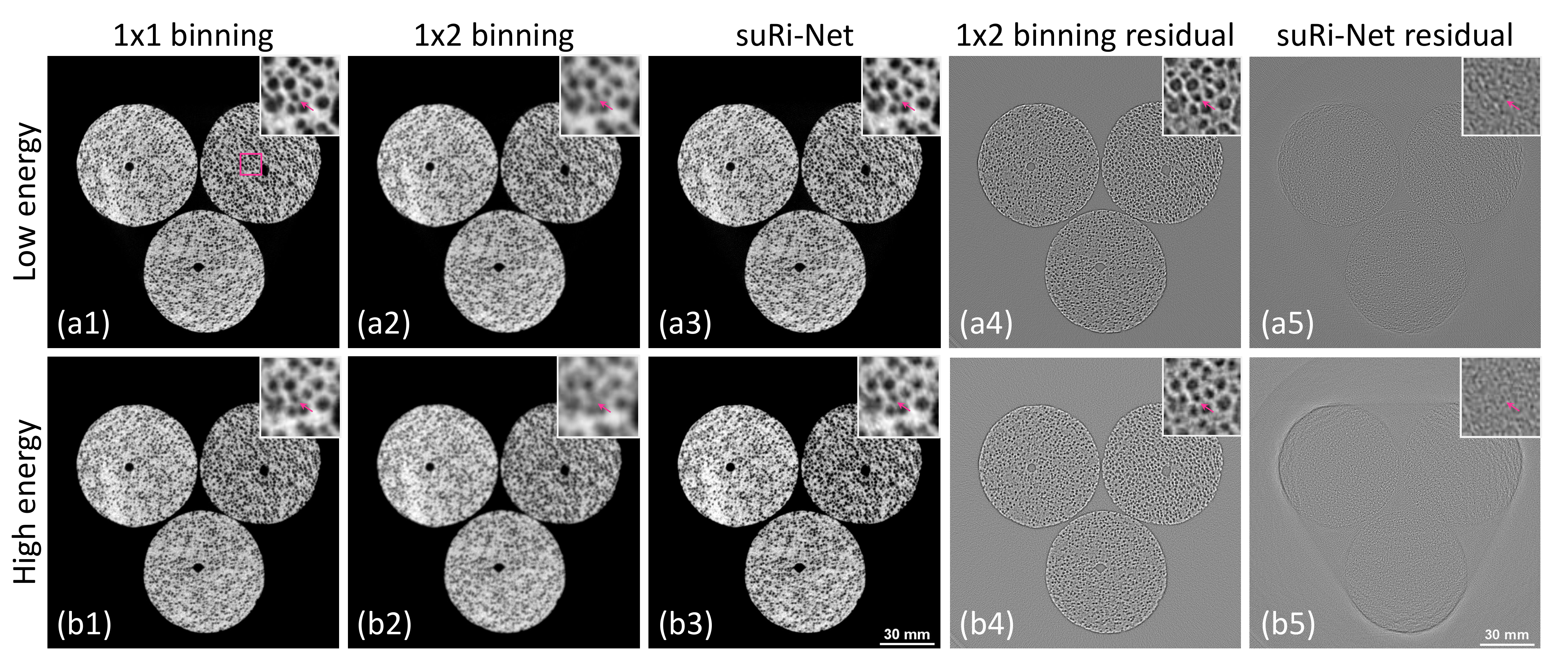}}
\caption{Imaging results of the limestone rods. The reference CT images obtained from 1x1 binning are depicted in the first column, the 1x2 binned CT images are depicted in the second column, and the results obtained from suRi-Net are depicted in the third column. The residual differences are depicted in the fourth and fifth columns. The display windows for the low-energy and high-energy CT images are [0.040, 0.295] cm$^{-1}$ and [0.054, 0.223] cm$^{-1}$, respectively. The display windows for the residual images are [-0.057, 0.060] cm$^{-1}$ and [-0.042, 0.033] cm$^{-1}$, respectively.}
\label{Limestone}
\end{figure*}

The CT imaging results of the porous limestone rods are shown in Fig.~\ref{Limestone}. The high contrast between air and limestone provides a great opportunity to investigate the  performance of suRi-Net in improving the CT image resolution. Specifically, high resolution CT images in the first column are the ground truth obtained from the $1\times1$ binned projections, low resolution CT images in the second column are obtained from the $1\times2$ binned projections, the suRi-Net reconstructed CT images are shown in the third column. The fourth and fifth columns present the residual images obtained by subtracting the $1\times2$ binning CT image and the suRi-Net reconstructed CT image from the $1\times1$ binning image, respectively. Compared to the CT images reconstructed from the $1\times2$ binned projections, clearly, the developed suRi-Net is able to significantly enhance the image resolution, especially in delineating the porous structures of the limestone rods. For both the top and bottom detector layers, similar performance are obtained. Since the scintillator material on the bottom layer is thicker than the top layer, therefore, the CT images obtained from the bottom layer look a little more blurry than the ones obtained from the top layer, see the images in the bottom row of Fig.~\ref{Limestone}.

\begin{figure*}[htb]
\centerline{\includegraphics[width=0.67\linewidth]{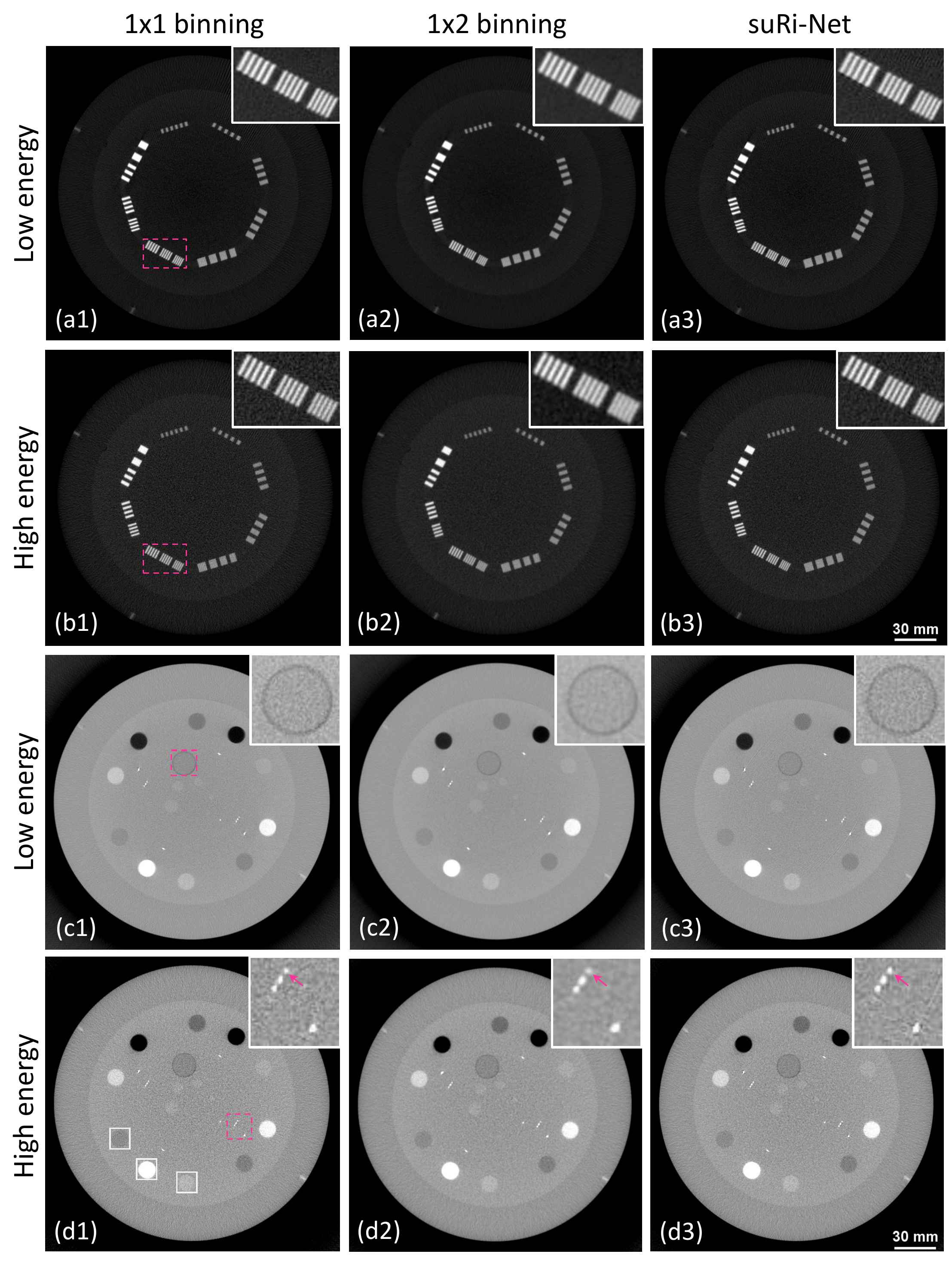}}
\caption{Imaging results of the Catphan phantom. The reference CT images obtained from 1x1 binning are depicted in the first column, the 1x2 binned CT images are depicted in the second column, and the results obtained from suRi-Net are depicted in the third column. From top to bottom, the display windows are [0.166, 0.467] cm$^{-1}$, [0.151, 0.428] cm$^{-1}$, [0.028, 0.302] cm$^{-1}$ and [0.073, 0.259] cm$^{-1}$, respectively.}
\label{Catphan_results}
\end{figure*}

The CT imaging results of the Catphan phantom are shown in Fig.\ref{Catphan_results}. Specifically, the CT images in the first and second rows correspond to the CTP714 module, and the CT images in the third and fourth rows correspond to the CTP682 module. According to the zoomed-in images (highlighted in pink rectangles), clearly, results generated from the suRi-Net in the third column show comparable spatial resolution performance as of the reference results obtained in the first column. For example, the highlighted high contrast bar patterns in Fig.\ref{Catphan_results}(a3) and (b3) are much sharpen than the ones in Fig.\ref{Catphan_results}(a2) and (b2). Similarly, the high contrast bearing balls in Fig.\ref{Catphan_results}(d3) are much less blurry than the one in Fig.\ref{Catphan_results}(d2). Again, the results of Catphan phantom demonstrate the capability of the proposed suRi-Net method is greatly enhancing the resolution of the CT images.

\begin{figure}[htb]
\centerline{\includegraphics[width=1.0\linewidth]{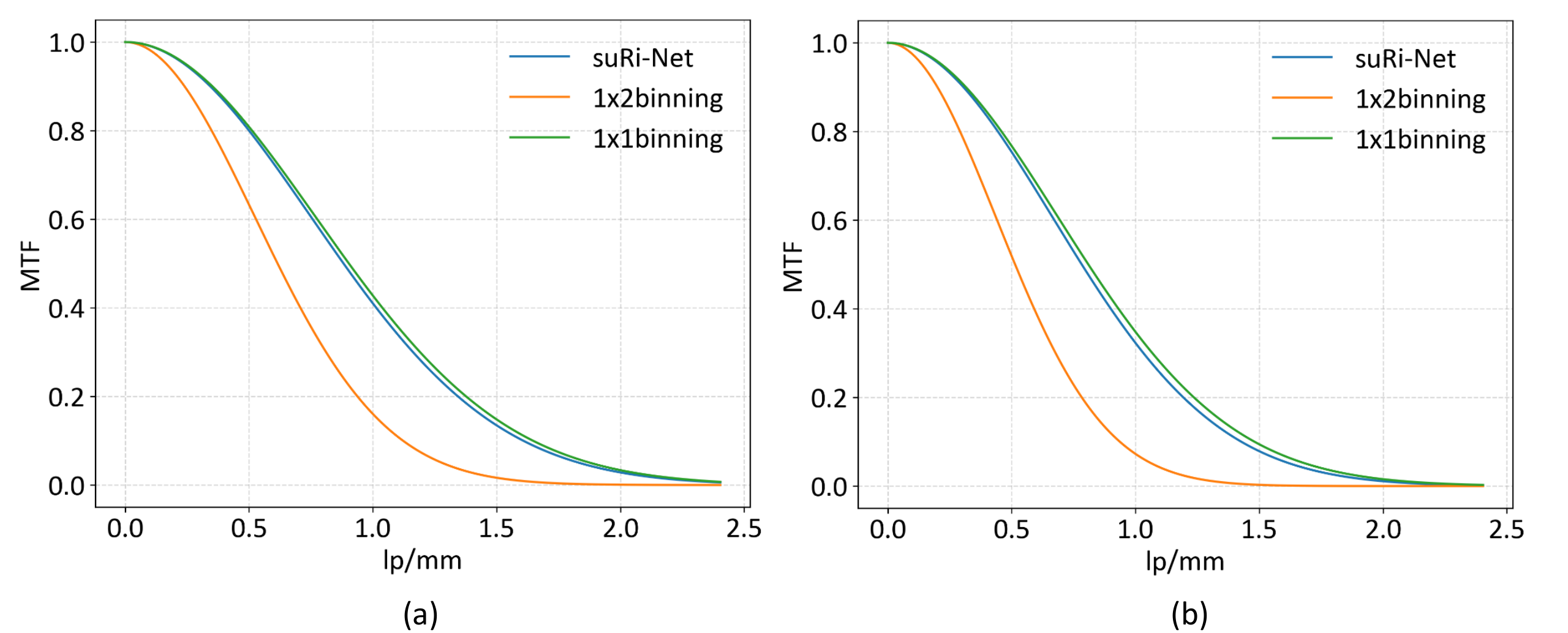}}
\caption{The measured MTF curves of the Teflon rod in CTP682 module.(a) MTF curves on the top layer. (b) MTF curves on the bottom layer.}
\label{MTF}
\end{figure}

The MTF curves measured\cite{Droege1984ModulationTF} on the Teflon rod (highlighted by white box in Fig.\ref{Catphan_results}(d1)) in CTP682 module are plotted in Fig.\ref{MTF}. As seen, the MTF curves obtained from suRi-Net are slightly lower than the ones obtained from the reference (1$\times$1 binning), but greatly higher than the ones obtained from the 1$\times$2 binned projections. Quantitatively, it is found that the spatial frequencies corresponding to $20\%$ MTFs in Fig.\ref{MTF}(a) on the low-energy CT image are 0.92 lp/mm, 1.34 lp/mm and 1.37 lp/mm for the 1$\times$2 binning, suRi-Net and 1$\times$1 binning, respectively. Similarly, the spatial frequencies corresponding to $20\%$ MTFs in Fig.\ref{MTF}(b) on the high-energy CT image are 0.77 lp/mm, 1.19 lp/mm and 1.22 lp/mm for the 1$\times$2 binning, suRi-Net and 1$\times$1 binning, respectively. Quantitatively, the suRi-Net improves the spatial resolution of the CT images by 45$\%$ and 54$\%$ in the top and bottom layer compared with the CT images reconstructed directly from the 1$\times$2 binned projections, demonstrating the strong capability of suRi-Net in enhancing the spatial resolution of DL-FPD based CBCT imaging. 

\begin{figure*}[htb]
\centerline{\includegraphics[width=1\linewidth]{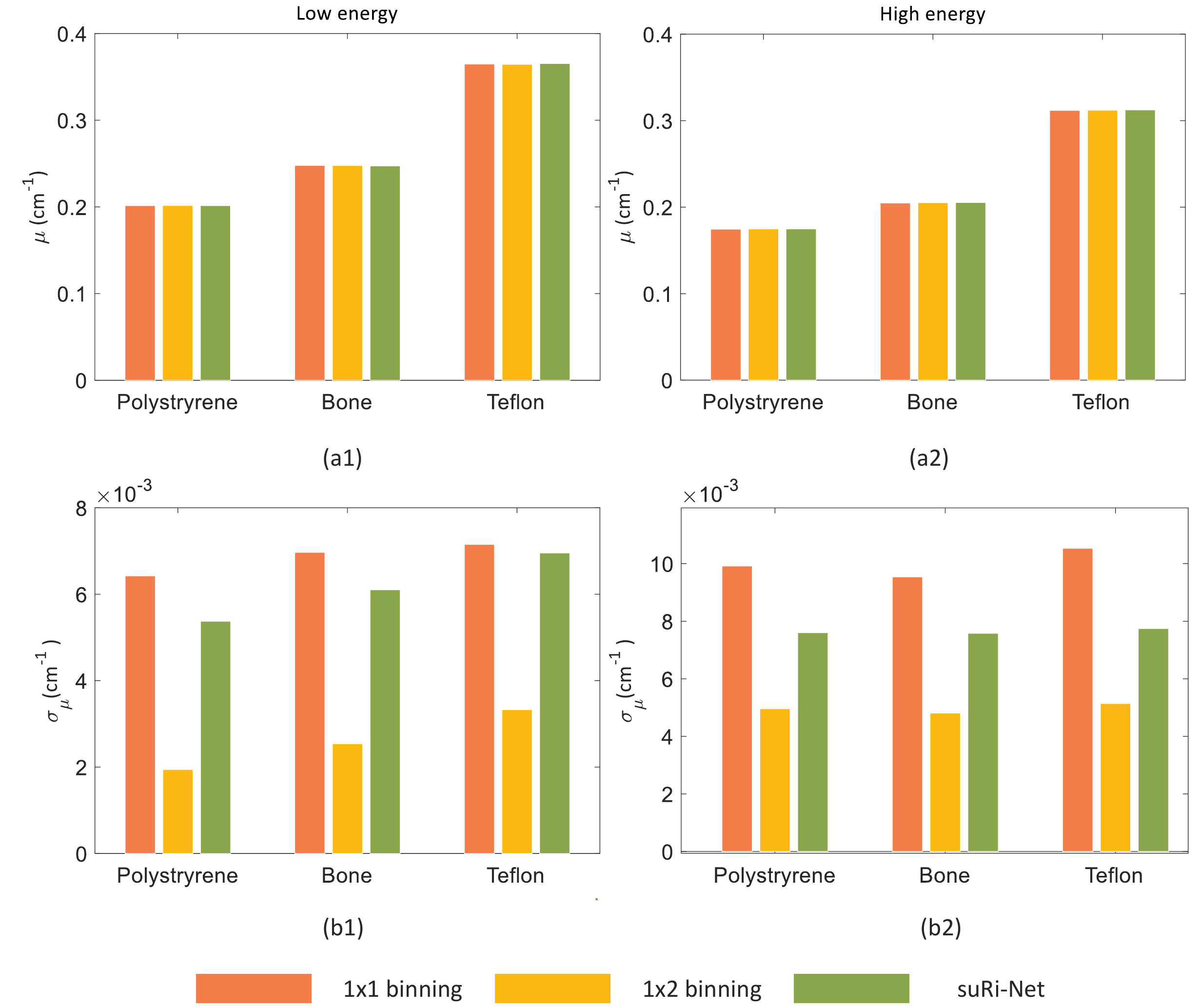}}
\caption{Bar plots of the mean and standard deviation of the measured signals for the Polystyrene, Bone, and Teflon materials in Catphan. Plots in the first column are obtained from the low-energy CT images, and plots in the second column are obtained from the high-energy CT images.}
\label{Catphan_mean_std}
\end{figure*}

In addition, the mean signal values and signal standard deviations of Polystyrene, Bone, and Teflon materials are compared, see Fig.\ref{Catphan_mean_std}. Clearly, suRi-Net can well maintain the signal values, and can also suppress the CT image noise (compared to the 1$\times$1 binned reference CT images). Due to the large detector binning, the image noise is much lower in the 1$\times$2 binned case. Furthermore, the noise levels of the high-energy CT images are higher than that of the low-energy CT images, see Fig.\ref{Catphan_mean_std}(b1) and (b2). This is because less photons are detected by the bottom layer during our experiments.

\begin{figure*}[htb]
\centerline{\includegraphics[width=0.95\linewidth]{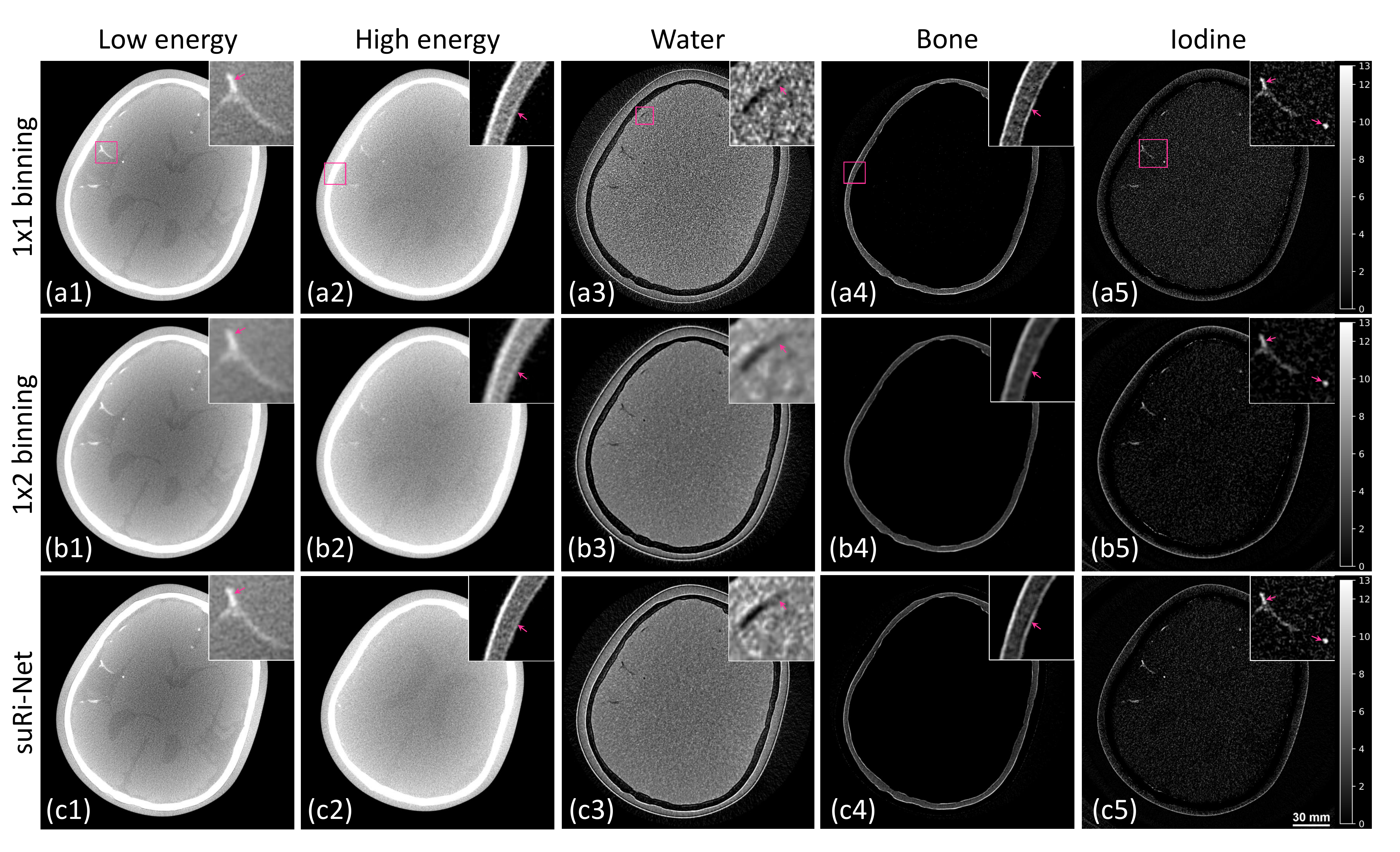}}
\caption{Imaging results of the KYOTO phantom. Results in the first and second columns show the dual-energy CT images. The display windows are [0.114, 0.206] cm$^{-1}$ and [0.088, 0.182] cm$^{-1}$, respectively. Images in the third and fifth columns are the decomposed water, bone and iodine density maps. The display windows are [0, 2.120] g/cm$^{3}$, [0.044, 5.557] g/cm$^{3}$ and [0, 13] mg/cm$^{3}$, respectively.}
\label{KYOTO_results}
\end{figure*}

Figure~\ref{KYOTO_results} shows the super-resolution CT imaging results of the KYOTO phantom. In it, the CT images obtained from the top and bottom detector layers are presented in the first and second columns, respectively, and the decomposed water, bone and iodine basis maps are presented in the third to fifth columns, respectively. Several typical anatomical structures are highlighted, from which the advancement of suRi-Net in generating high resolution CT images can be clearly demonstrated. For instance, the iodine contrast-enhanced vessels and skull reconstructed from suRi-Net in Fig.~\ref{KYOTO_results} (c1) and (c2) are much sharper than the ones reconstructed directly from the 1$\times$2 binned projections in Fig.~\ref{KYOTO_results} (b1) and (b2). Because suRi-Net is able to generate high resolution CT images for both low-energy and high-energy, therefore, the resultant decomposed basis maps from suRi-Net also show superior spatial resolution than the ones obtained from the 1$\times$2 binned dataset, see the arrow pointed regions on Fig.~\ref{KYOTO_results} (c3), (c4) and (c5).



\section{DISCUSSIONS AND CONCLUSION}\label{sec:conclusion}
In this study, a novel super resolution CBCT imaging method is developed by taking the advantages of the intrinsic detector element offsets (assuming inclined fan-beam incidence) between the dual-layers. Such intrinsic detector element offsets help to increase the spatial sampling density, and thus enable super resolution CBCT imaging. A simple mathematical model is assumed and derived to explain the mechanism of recovering the high resolution spatial information. Within the imaging model, the sub-pixels either on the top detector layer or on the bottom detector layer are categorized into two groups. Driven by such a mathematical model, a deep learning RNN framework is developed to extract the high resolution spatial information. Results demonstrate that CBCT images reconstructed from the suRi-Net show similar spatial resolution as of the ones acquired with small detector element binning. Quantitatively, this new method is able to improve the spatial resolution of CBCT images by over 45.0$\%$ (measured with the spatial frequency at 20\% MTF). In addition, the decomposed basis images also maintain high signal accuracy.

Instead of extracting the super resolution information purely relying upon the deep learning techniques, suRi-Net retrieves such high resolution information more heavily upon the underlying imaging physics of DL-FPD. In fact, the spatial resolution information that is finer than the native detector element dimension has already been acquired and encoded in DL-FPD, see Fig.~\ref{Imaging_model}. As discussed previously, such super resolution information may be obtained analytically via a certain mathematical model. Due to the non-linear responses in the top and bottom detector layers, however, it is quite challenging to implement such analytical solution in practice. As a consequence, the RNN network is employed in this work to retrieve the super resolution information from the non-linear responses of the top and bottom detector layers. In particular, certain paired detector responses of the top and bottom layers are fed into the corresponding RNN layer together with the estimated responses from the previous RNN layer to generate the high spatial resolution responses. By repeating such procedure across the entire detector plane, high resolution projections can be eventually obtained.

The assumption of equal sub-pixel dimension ($\delta_k = 0.5\delta_{del}$) made in this study is very empirical. Such an average is statistically viable, see Fig.~\ref{subpixel_size}(b). With this assumption, the sub-pixel detector response in any $\delta_k$ (having varied physical widths) can be easily represented in a computer, which by default assumes equal sub-pixel dimension in image storage. Regardless, this simple assumption may set an upper limit to the resolution of the restored projections. This interesting topic is beyond the scope of this study, and would be investigated as a future work. In addition, the gap between the top and bottom detector layers needs to be moderate. If the gap is too small, the proposed suRi-Net method might become invalid. For example, if the top and bottom detector layers are contacted, namely, $\delta'_{del}\approx\delta_{del}$, the intrinsic detector element shifts $\delta_k$ would be tiny (close to 0) across the entire detector plane. To facilitate the use of the proposed suRi-Net method, as a consequence, moderate gap between the top and bottom detector layers is needed. In practice, this is not difficult to achieve when fabricating the DL-FPD.

This study has several limitations. First, the CBCT images acquired from the two detector layers have inconsistent spatial resolutions due to their varied scintillator thicknesses. This problem could be mitigated by making the DL-FPD with equal scintillator thickness\cite{Gu2022Evaluation}. Second, Compton scattering was neglected by assuming a narrow beam (15.0 mm) width. For CBCT imaging with full field-of-view, Compton scatters should be properly corrected\cite{Sisniega2015HighfidelityAC}. Third, super resolution CBCT imaging is only verified on the axial plane, and high resolution CBCT imaging along both the horizontal and vertical directions have not been investigated yet. It is believed that the proposed suRi-Net is able to complete such super resolution imaging task by processing the low-energy and high-energy projections for two times: first along the horizontal direction, and then along the vertical direction. Fourth, the feasibility of using suRi-Net in dynamic CBCT imaging would be investigated in future. In this case, the spatial and temporal resolution of the four-dimensional (4D) CBCT imaging may be significantly enhanced.

In conclusion, a super resolution CBCT imaging method, named as suRi-Net, is developed with DL-FPD by using of the intrinsic sub-pixel shifts between the top and bottom detector layers. A dedicated RNN network is designed to retrieve such high resolution signal responses. Experimental results demonstrate that suRi-Net can improve the image spatial resolution by up to 45\%. In future, suRi-Net may greatly improve the image resolution of dual-energy CBCT with DL-FPD.


\bibliography{mybib}
\bibliographystyle{unsrt}

\end{document}